\documentclass[a4paper,conference]{IEEEtran}

\usepackage{cite}      
\usepackage{graphicx}  
\usepackage{psfrag}                            
\usepackage{subfigure} 
\usepackage{url}      
\usepackage{stfloats}                          
\usepackage{amsmath}   
\usepackage{array}
\usepackage{fancyhdr}

\begin{document}

\title{Regions of multistability in some 
low-dimensional logistic models with excitation type coupling}

\author{\authorblockN{Ricardo L\'opez-Ruiz}
\authorblockA{Facultad de Ciencias \\
Universidad de Zaragoza \\ E-50009 Zaragoza, Spain\\
Email: rilopez@unizar.es}
\and
\authorblockN{Dani\`ele Fournier-Prunaret}
\authorblockA{SYD and LESIA \\
Institut National des Sciences Appliqu\`ees \\
31077 Toulouse Cedex, France\\
Email: Daniele.Fournier@insa-toulouse.fr}}

\maketitle

%\thispagestyle{fancy}
%\markboth{\scriptsize IEEE Workshop on Nonlinear Maps and Applications (NOMA'07)}
%{\scriptsize LATTIS-INSA, Toulouse University, France, Dec. 13-14, 2007}

%***************************************

\begin{abstract}

A naive model of many networked logistic maps with an excitation type coupling
[Neural Networks, vol. 20, 102--108 (2007)], which is an extension of other
low dimensional models, has been recently proposed
to mimic the waking-sleeping bistability found in brain systems.
Although the dynamics of large and complex aggregates of elementary 
components can not be understood nor extrapolated from the properties of 
a few components, some patterns of behavior could be conserved independently 
of the topology and of the number of coupled units.
Following this insight, we have collected several of those systems where a few 
logistic maps are coupled under a similar mutual excitation scheme.
The regions of bi- and multistability of these systems
are sketched and reported.

\end{abstract}

\IEEEpeerreviewmaketitle

\section{Introduction}

The brain is a natural networked system \cite{cajal}. 
The understanding of this complex system
is one of the most fascinating scientific tasks today, 
concretely how this set of millions of neurons can interact among them to give 
rise to the collective phenomenon of human thinking \cite{sfn}, or, 
in a simpler and more realistic approach, what neural features can make 
possible, for example, the birdsongs \cite{mindlin}.   

\begin{figure}[h]
 \center\includegraphics[height=.15\textheight]{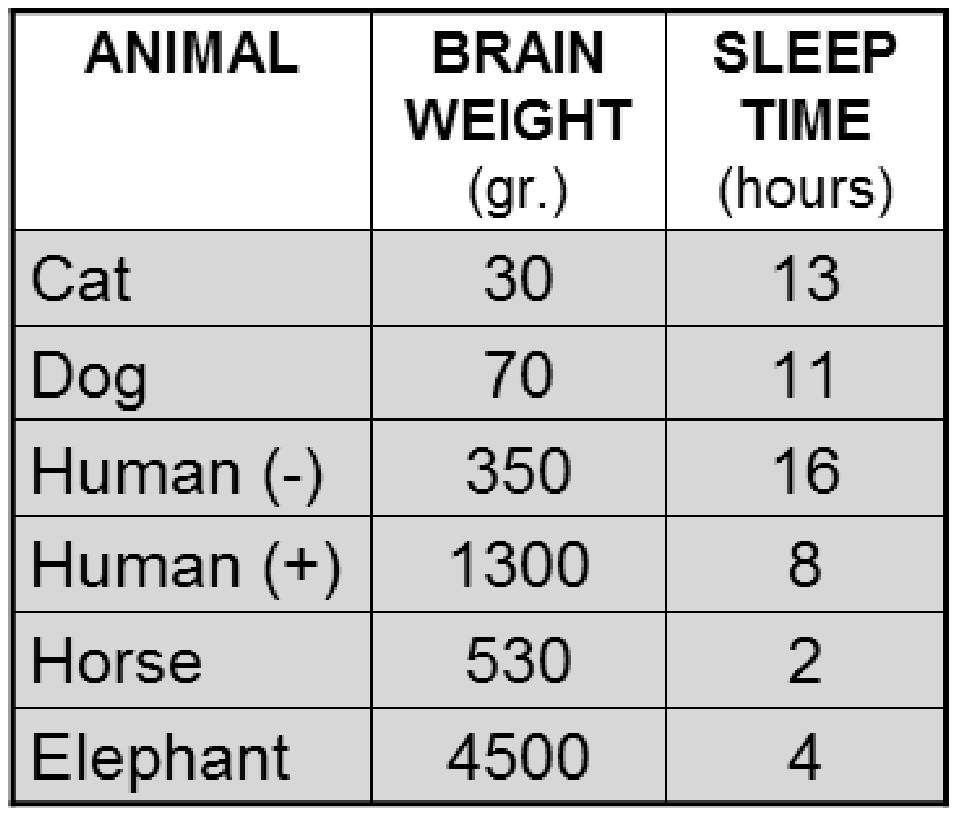}
 \label{hola}
  \caption{Brain size and mean sleeping time for different animals. (Humans(-) represent
  new born humans and Humans(+) represent middle age humans).}
\end{figure}

\begin{figure}[h]
 \center\includegraphics[height=.2\textheight]{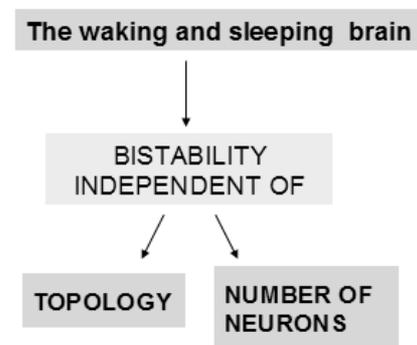}
  \caption{The waking-sleeping brain.}
  \label{waksleep}
\end{figure}

What is clear is that many of the brain functions are not dependent 
neither of its architecture nor of its size. Take, for instance, 
the universal property in mammals and birds of the sleep-wake cycle \cite{winfree,bar-yam},
a regular daily behavior that is closely synchronized with the
cycle of sun light. It is observed that, in general, 
large animals tend to sleep less than small animals (some sleeping time data are 
shown in Fig. 1).
Hence, the bistable sleep-wake behavior is one of the emergent attributes that does not depend 
on the precise architecture of the brain nor on its size \cite{lopezruiz07,lopezruiz007}. 
If we represent the brain as a complex network this property would mean that this possible bistability where 
large groups of neurons can show some kind of synchronization should not 
depend on the topology (structure) nor on the number of nodes (size) of the network 
(Fig. \ref{waksleep}).

Then it seems that it is essential the type of local dynamics and coupling among nodes 
that must be implemented in order to get bistability as the possible dynamical panorama
of a complex network. In the next section, 
we give some possible (excitation/inhibition) strategies for the coupling and 
the local (logistic) dynamics which should be implemented in a few or many units network 
in order to find bistable behavior between two complete synchronized states \cite{lopezruiz007}. 
In Section 3, some discrete logistic models in two
and three dimensions, showing bi- and multistability under the excitation scheme, are presented. 
Finally, the last section contains our conclusions.

\section{General Model}

\begin{figure}[h]
 \center\includegraphics[height=.05\textheight]{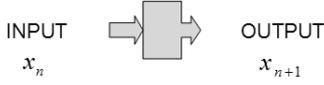}
  \caption{Discrete nonlinear model for the local evolution of a functional unit.}
  \label{1-osc}
\end{figure}

Our approach consider the so called 
{\it functional unit}, i.e. a neuron or group of neurons, as a discrete 
nonlinear oscillator \cite{kuhn} with two possible states: 
active (meaning one type of activity) 
or not (meaning other type of activity).
Hence, in this naive vision of the brain as a networked system, 
if $x_n^i$, with $0<x_n^i<1$, represents a measurement of the $ith$ 
functional unit activity at time $n$, 
it can be reasonable to take the most elemental local nonlinearity, 
for instance, a logistic evolution \cite{may,mira}, which presents a quadratic term, 
as a first toy-model for the local neuronal activity (Fig. \ref{1-osc}):

\begin{equation}
x^i_{n+1} = \bar p_i\;x^i_n(1-x^i_n).
\label{eq0}
\end{equation}

It presents only one stable state for each $\bar p_i$. Then, there is no
bistability in the basic component of our models. For $\bar p_i<1$,
the dynamics dissipates to zero, $x_n^i=0$, then it can represent the
functional unit with no activity. For $1<\bar p_i<4$, the dynamics is
non null and it would represent an active functional unit. 

We can suppose that this local parameter $\bar p_i$ is controlled by the signals of 
neighbor units, simulating in some way the effect of the synapses among neurons.
Excitatory and inhibitory synaptic 
couplings have been shown to be determinant on the synchronization of neuronal firing. 
For instance, facilitatory connections are important to explain the neural 
mechanisms that make possible the object representation by synchronization
in the visual cortex \cite{eckhorn}.
While it seems clear that excitatory coupling can lead to synchronization, 
frequently inhibition rather than excitation synchronizes 
firing \cite{abbott}. The importance of these two kinds of coupling mechanisms
has also been studied for other types of neurons, v.g., motor neurons \cite{koening}.

If a neuron unit simultaneously processes a plurality of binary input signals,
we can think that this local information processing is reflected by the 
parameter $\bar p_i$. The functional
dependence of this local coupling on the neighbor states is essential
in order to get a good brain-like behavior (i.e., as far as the
bistability of the sleep-wake cycle is concerned) of the network. 
As a first approach, we can take $\bar p_i$ as a linear 
function depending on the actual mean value, $X_n^i$, of the neighboring 
signal activity and expanding the interval $(1,4)$ in the form:
\begin{eqnarray}
\bar p_i & = & p_i\;(3X_n^i+1), \;\;\;\;\;\;\; (excitation \;\; coupling) \label{eq1}\\
& {or} & \nonumber \\
\bar p_i & = & p_i\;(-3X_n^i+4), \;\;\;\; (inhibition \;\; coupling) \label{eq11}
\end{eqnarray}
with 
\begin{equation}
X_n^i={1\over N_i}\sum_{j=1}^{N_i}x_n^j.
\label{eq2}
\end{equation}
$N_i$ is the number of neighbors of the $ith$ functional unit, and
$p_i$, which gives us an idea of the interaction of the functional unit
with its first-neighbor functional units, is the control parameter.
This parameter runs in the range $0<p_i<p_{max}$, where $p_{max}\succeq
1$. When $p_i=p$ for all $i$, the dynamical behavior of these networks 
with the excitation type coupling \cite{lopezruiz07} presents an attractive global
null configuration that has been identified as the {\it turned off}
state of the network.  Also they show a completely synchronized
non-null stable configuration that represents the {\it turned on}
state of the network. Moreover, a robust bistability between these two perfect 
synchronized states is found in that particular model 
(see \cite{lopezruiz07} and \cite{oprisan09} for more details). 
For different models with a few
coupled functional units we sketch in the next subsections the regions 
where they present a multistable behavior.

\section{Models with Excitation Coupling}

\subsection{Model of two functional units}

The case of two interconnected $(x_n,y_n)$
functional units \cite{lopezruiz04} 
under the {\it(excitation, excitation)} 
scheme is given by the coupled equations:
\begin{eqnarray}
x_{n+1} & = & p \;(3y_n+1)x_n(1-x_n),\\
y_{n+1} & = & p \;(3x_n+1)y_n(1-y_n).
\label{2-osc}
\end{eqnarray}
The regions of the parameter space 
where we can find bistability was presented in \cite{lopezruiz007}.

\subsection{Models of three functional units}

\subsubsection{Model with local mutual excitation}

The case of three alternatively interconnected $(x_n,y_n,z_n)$
functional units \cite{fournier06} under a mutual excitation scheme
is given by the coupled equations:
\begin{eqnarray}
x_{n+1} & = & p \;(3y_n+1)x_n(1-x_n), \\
y_{n+1} & = & p \;(3z_n+1)y_n(1-y_n), \\
z_{n+1} & = & p \;(3x_n+1)z_n(1-z_n).
\label{3-osc+}
\end{eqnarray}

The regions of the parameter space 
where bistability can be found are sketched in \cite{lopezruiz007}.

\subsubsection{Model with global mutual excitation}

The case of three globally interconnected $(x_n,y_n,z_n)$
functional units \cite{fournier06} under a mutual excitation scheme 
is given by the coupled equations:
\begin{eqnarray}
x_{n+1} & = & p\; (x_n+y_n+z_n+1)x_n(1-x_n), \\
y_{n+1} & = & p\; (x_n+y_n+z_n+1)y_n(1-y_n), \\
z_{n+1} & = & p\; (x_n+y_n+z_n+1)z_n(1-z_n).
\label{3-osc-}
\end{eqnarray}
For the whole range of the parameter, $0<p<1.17$, bistability is present 
in this system \cite{lopezruiz007}.

\subsubsection{Model with partial mutual excitation}

\begin{figure}[h]
 \center\includegraphics[height=.1\textheight]{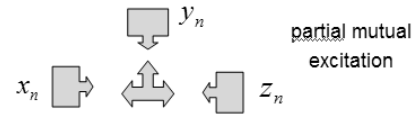}
  \caption{Three partially coupled functional units under the excitation scheme.}
  \label{fig-3}
\end{figure}

The new case of three partially interconnected $(x_n,y_n,z_n)$
functional units under a mutual excitation scheme is represented in Fig. \ref{fig-3}. 

The dynamics of the system is given by the coupled equations:
\begin{eqnarray}
x_{n+1} & = & p\; (3(y_n+z_n)/2+1)x_n(1-x_n), \\
y_{n+1} & = & p\; (3(x_n+z_n)/2+1)y_n(1-y_n), \\
z_{n+1} & = & p\; (3(x_n+y_n)/2+1)z_n(1-z_n).
\label{3-osc-1}
\end{eqnarray}

\begin{figure}[h] 
 \center\includegraphics[height=.28\textheight]{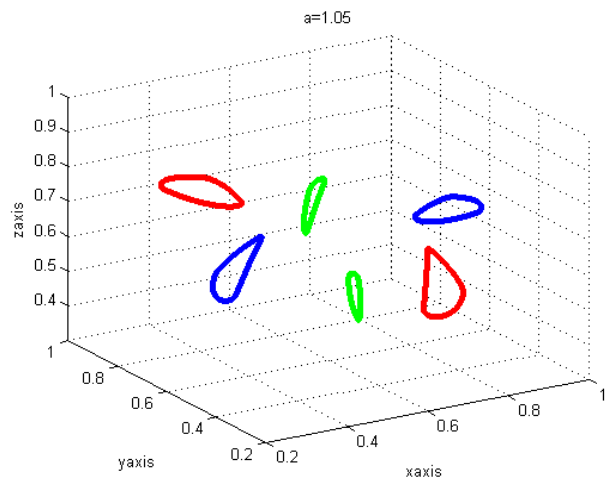}
  \caption{Multistability in $3$ functional units with partial excitation type coupling.
  The system presents three order-$2$ ICC for $p=a=1.05$.}
  \label{fig6}
\end{figure}

\begin{figure}[h]
 \center\includegraphics[height=.28\textheight]{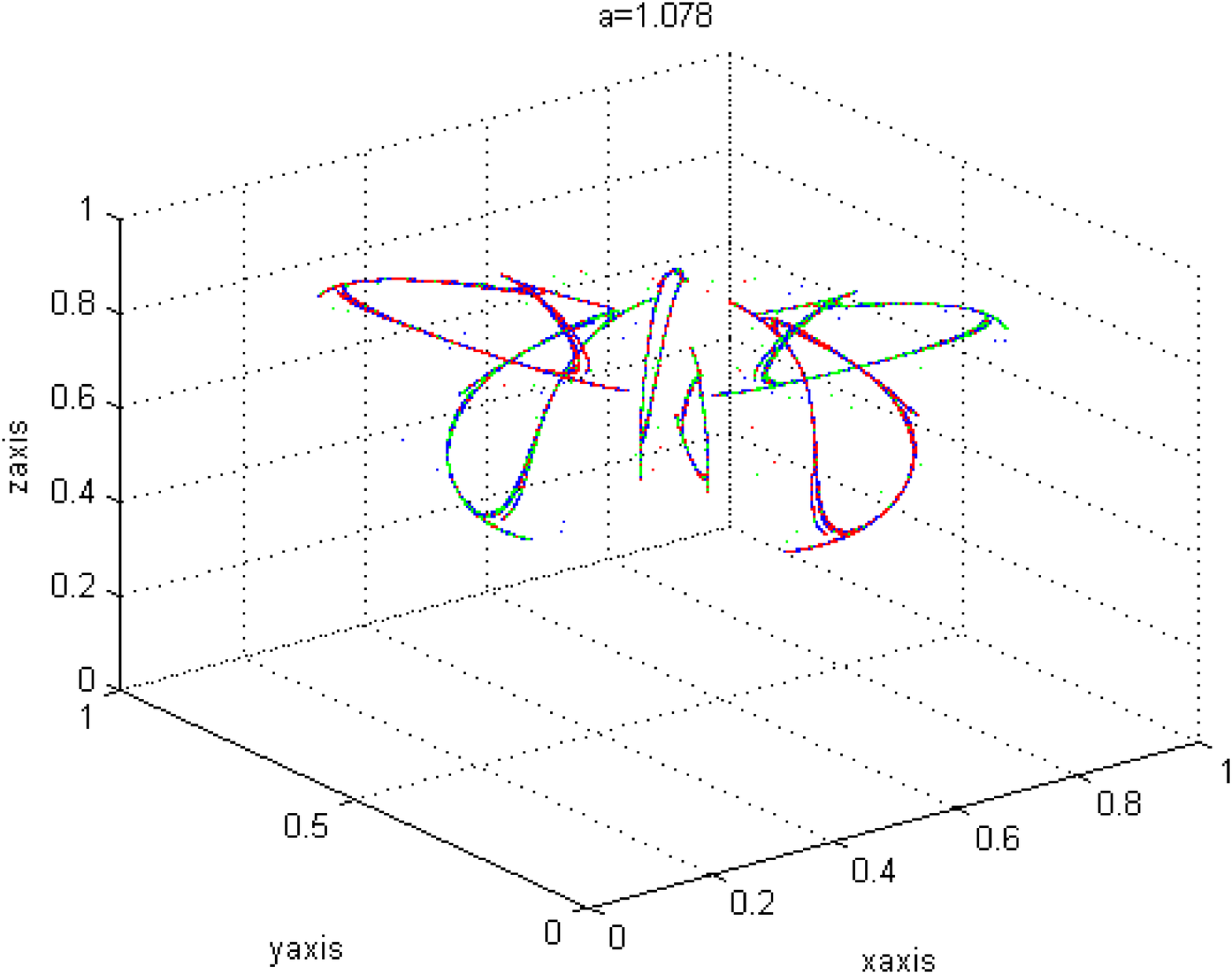}
  \caption{Multistability of three chaotic cyclic attractors of order $2$ for $p=a=1.078$.}
  \label{fig7}
\end{figure}

\begin{figure}[h]
 \center\includegraphics[height=.28\textheight]{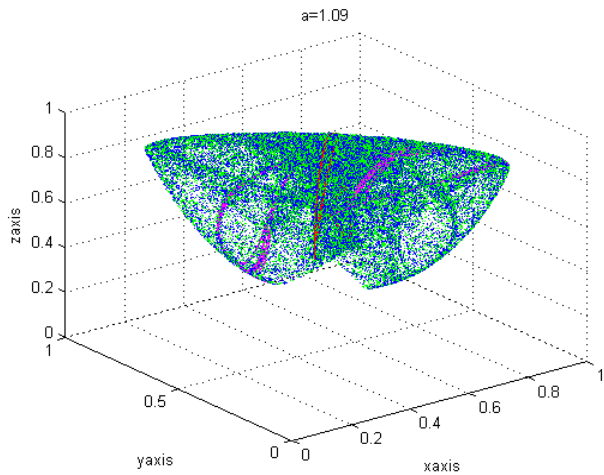}
  \caption{An unique chaotic attractor for $p=a=1.09$.}
  \label{fig8}
\end{figure}

The rough inspection of this system puts in evidence 
the existence of different regions of multistability 
in the parameter space. These are:

\begin{itemize}
\item For $0.93<p<1.04$, there are coexistence among
three cycles of period-$2$.
\item For $1.04<p<1.06$, the three cycles bifurcate giving rise
to three order-$2$ ICC (Fig. \ref{fig6}).
\item For $1.06<p<1.08$, the system can present three mode-locked periodic orbits or three
chaotic cyclic attractors, each one with period multiple of $6$, or three chaotic cyclic attractors 
of order $2$ (Fig. \ref{fig7}). 
\item For $p>1.08$, the chaotic cyclic attractors collapse in an unique chaotic attractor (Fig. \ref{fig8}).
\end{itemize}
Also, other bistable and multistable situations can be found for some particular values of the parameter $p$
in the former intervals.

\section{Conclusions}

Different neural systems can exhibit similar dynamical properties 
despite having different architectures, different sizes or different 
complexity \cite{varona}. To be able of reproducing, even qualitatively,
some common features observed in those systems could be considered 
as an effective advance. Thus, specifically, if the sleep-wake cycle is 
interpreted as a bistability in the global behavior of a neural system,
it might be of some interest to dispose of a model that reproduces 
this type of phenomenology.
In this presentation, different coupling schemes for networks with local logistic
dynamics are proposed. It is observed that this kind of couplings
generates a global bistability between two different dynamical states.
This property seems to be topology and size independent. This is a direct
consequence of the local mean-field multiplicative coupling among the
first-neighbors. 
Following this insight, different low-dimensional systems with logistic components  
coupled under an excitation scheme have been collected, and the regions where 
the dynamics shows (multi-) bistability have been identified.

%%%%%%%%%%%%%%%%%%%%%%%%%%%%%%%%%%%%%%%%%%%%
%% BACKMATTER
%%%%%%%%%%%%%%%%%%%%%%%%%%%%%%%%%%%%%%%%%%%%

%\newpage

\end{document}